\documentclass[final]{revtex4} 
\usepackage{layout}
\usepackage{amsmath}
\usepackage{textcomp}
\usepackage{hyperref}
\begin{document}
\title{Fluctuation of gauge field for general nonlinear Fokker-Planck 
equation and covariant version of Fisher information matrix}
\author{{\sc Takuya Yamano}}
\email[Email: ]{yamano@amy.hi-ho.ne.jp}
\affiliation{
Department of Mathematics and Physics, Faculty of Science, 
Kanagawa University,\\
2946, 6-233 Tsuchiya, Hiratsuka, Kanagawa 259-1293, Japan}

\begin{abstract}
We clarify a strong link between general nonlinear Fokker-Planck equations with 
gauge fields associated with nonequilibrium dynamics and the Fisher information of 
the system. The notion of Abelian gauge theory for the non-equilibrium Fokker-Planck 
equation has proposed in the literature, in which the associated curvature represents 
internal geometry. We present the fluctuation of the gauge field can be decomposed 
into three parts. We further show that if we define the Fisher information matrix 
by using a covariant derivative then it gives correlation of the flux components 
but it is not gauge invariant.
\end{abstract}


\maketitle
\section{Introduction}
The phenomena pertinent to non-stationary dynamics are ubiquitous in nature. 
They are usually described by the time evolution of the probability distribution 
of the system. Among them, various types of the nonlinear Fokker-Planck equation 
(FPE) have been attracted much attention for analyzing anomalous diffusion 
processes found in the motion of particles in porous media and many other 
complex systems \cite{Frank}. \\

Recently, on the one hand, Feng and Wang \cite{Feng} introduced the concept of 
Abelian gauge field \cite{Peskin} for the linear FPE to legitimately understand 
the degree of non-stationarity of systems by attributing the non-vanishing 
curvature caused by the field. The first aim of this paper is to advance the 
horizon of this approach in the context of the general nonlinear FPE. To be more 
precise, we focus on how the fluctuation of the gauge field is understood by 
information related quantity i.e., the Fisher information and its matrix 
\cite{Cover}.\\

In some sense, the Fisher information can be regarded as a quantity that lies behind 
physical laws \cite{Frieden} and it involves the spatial gradient of the distribution 
function. Furthermore, this quantity plays a role of an upper bound on the entropy 
production (e.g. \cite{Brody,TY1} and references therein).
Since the covariant derivative shifts the gradient by the amount of the gauge field, 
when it exists, this effect should naturally modify the information. However, so far, 
there is no attempt to understand the properties of the covariant versions of the 
Fisher information matrix. This paper considers this gap.\\

The paper is organized as follows: in Sect.~2 we fix the notations of our focused 
general nonlinear FPE. In Sect.~3, we give the precise formulation of the fluctuation 
of the gauge field. Sect.~4 considers the possible alternatives for the Fisher 
information matrix to understand nonequilibrium systems. We introduce the covariant 
derivative associated with the gauge field and explore how it defines the Fisher 
information matrix by applying to the stationary state distribution. In the last 
section, we restate the final conclusions.

\section{A general nonlinear Fokker-Planck equation and its gauge field}
The general nonlinear Fokker-Planck equations (FPEs) we deal with are of the 
form \cite{Frank},
\begin{eqnarray}
\frac{\partial P(\vec{x},t)}{\partial t}=&-&\sum_i\frac{\partial}{\partial x_i}
[\mathcal{F}_i(\vec{x})K(P(\vec{x},t))]\nonumber\\
&+&\sum_{ij}\frac{\partial^2}{\partial x_i\partial x_j}[D_{ij}(\vec{x})
G(P(\vec{x},t))]P(\vec{x},t),
\end{eqnarray}
where $\mathcal{F}_i(\vec{x})$ is the $i$th component of the external force and 
$D_{ij}(\vec{x})$ is a time independent diffusion tensor which takes positive value. 
When the functions take $K(P)=P$ and $G(P)=1$, the above equation becomes the linear 
FPE. Since this form encompasses wide variety of nonlinear FPEs, the investigation of 
some universal features may be useful. As one of these studies, we mention that the 
entropy production under this equation was studied recently in Ref. \cite{Casas}. 
In these situations, however, considerations based on the concept of the gauge field 
are still missing. This fact comprises our first motivation in the present section. 
In terms of the probability current, 
\begin{eqnarray}
\mathsf{j}_i(\vec{x})=\mathcal{F}_i(\vec{x})K(P(\vec{x},t))-
\sum_j\frac{\partial}{\partial x_j}[D_{ij}(\vec{x})G(P(\vec{x},t))]P(\vec{x},t)
\end{eqnarray}
we can rewrite the above equation as the continuity equation 
$\partial_t P(\vec{x},t)+\textrm{div} \mathsf{j}(\vec{x})=0$. 
In line with Ref. \cite{Feng}, we introduce a covariant derivative  
$\nabla_i=\partial_{x_i}+A_i$, where $A_i$ is the $i$th component of the gauge field 
associated with probability flux and the gradient of the distribution. Usually, 
the covariant derivative operates on a vector field, but it is also valid for a 
scalar function (i.e., the probability distribution in our case. See Appendix) similar 
to a wave function used in gauge theory in quantum mechanics. By choosing $A_i$ as 
\begin{eqnarray}
A_i=D_{ij}^{-1}(\vec{x})\left[ \frac{F_j(\vec{x})K(P)}{P}+
G\frac{\partial D_{ij}(\vec{x})}{\partial x_i}\right]+
\frac{\partial G(P)}{\partial x_i}+\left( G(P)-1\right)
\frac{\partial \ln(P)}{\partial x_i},
\end{eqnarray}
we find that the general nonlinear FPE can be expressed as the form of Fick's 
first law
\begin{eqnarray}
-D_{ij}(\vec{x})\nabla_j P=\mathsf{j}_i(\vec{x}).\label{eqn:gnFPE}
\end{eqnarray}
Here and hereafter, we use the summation convention for the repeated index $j$ 
to avoid the cumbersome expression. As pointed out in \cite{Feng}, the gauge 
field $A_i$ is a measure of the nonequilibrium dynamics in that the non-flat 
curvature $\partial_{x_i}A_j-\partial_{x_j} A_i\neq 0$ indicates the non-vanishing 
stationary flux $\mathsf{j}^{ss}(\vec{x})$. With this setting, we consider what 
structure we obtain for the picture of the gauge field.
\section{Fluctuation of the gauge field}
We shall concern with how the fluctuation of $A_i$ is related to the Fisher 
information. First, when the stationary state has the component-wise vanishing 
flux $\mathsf{j}_i^{ss}(\vec{x})=0$, we find that $A_i$ becomes a pure gradient 
from Eq.(\ref{eqn:gnFPE}) and we put $\langle A_i\rangle_{0}$ to denote the average 
of it over $P^{ss}$:
\begin{eqnarray}
\langle A_i\rangle_0=-\int \partial_{x_i}P^{ss}d\vec{x}.
\end{eqnarray}
We leave out the argument of $P^{ss}(\vec{x})$ for simplicity and the integration 
means $\int (\cdot) d\vec{x}=\int\cdots \int_{\mathcal{R}^n} (\cdot) dx_1\cdots dx_n$ 
on the whole $\mathcal{R}^n$ with $n$ being the number of the variables.
Similarly, for $\langle A_i^2\rangle$ we have 
\begin{eqnarray}
\langle A_i^2\rangle_0=\int \frac{(\partial_{x_i}P^{ss})^2}{P^{ss}}d\vec{x}.
\end{eqnarray}
This equals just to the Fisher information $I(P^{ss})$ of the system at stationary state. 
This is an important observation that links the fluctuation of the gauge field and the 
information conception. In general nonequilibrium states represented by the probability 
distribution $P$, $\mathsf{j}_i(\vec{x})$ is neither zero nor divergence free. Therefore, 
we have from Eq.(\ref{eqn:gnFPE}),
\begin{eqnarray}
A_i=-D_{ji}^{-1}(\vec{x})\frac{\mathsf{j}_j(\vec{x})}{P}
-\partial_{x_i}\ln P .
\end{eqnarray}
We then have for the average 
\begin{eqnarray}
\langle A_i\rangle_{\mathsf{j}}=-\int D_{ji}^{-1}(\vec{x}) \mathsf{j}_j(\vec{x})d\vec{x}
+\langle A_i\rangle_0,
\end{eqnarray}
where $\langle \rangle_{\mathsf{j}}$ denotes the average with respect to $P$, and
\begin{eqnarray}
\langle A_i^2\rangle_{\mathsf{j}}&=&\Big\langle\left(\frac{\partial_{x_i}P}{P}\right)^2\Big\rangle
+2\int D_{ji}^{-1}(\vec{x})\mathsf{j}_j(\vec{x})\frac{\partial_{x_i}P}{P}d\vec{x}+
\Big\langle D_{ji}^{-2}(\vec{x})\left(\frac{\mathsf{j}_j(\vec{x})}
{P}\right)^2\Big\rangle\nonumber\\
&=& I(P)+ 2\Big\langle D_{ji}^{-1}(\vec{x})v_j(\vec{x})\partial_{x_i}(\ln P)\Big\rangle
+\langle D_{ji}^{-2}(\vec{x}) v_j^2(\vec{x})\rangle,
\end{eqnarray}
where $v_i(\vec{x})\equiv \mathsf{j}_i(\vec{x})/P(\vec{x})$ is the probability flux velocity 
at stationary state. We omit the suffix $\mathsf{j}$ in the brackets at the right-hand side 
here and below when it is evident from the context. We redefine the average 
$\langle A_i \rangle_{\mathsf{j}}$ as the one measured from $\langle A_i \rangle_0$ i.e., 
$\langle A_i \rangle_{\mathsf{j}} - \langle A_i \rangle_0$. The response coefficient \cite{Res} 
may accordingly be defined as 
\begin{eqnarray}
\chi_{\mu \nu}=\frac{\langle A_\mu \rangle_{\mathsf{j}}-
\langle A_\mu \rangle_0}{\mathsf{j}_\nu}\Big|_{\mathsf{j}=0}.
\end{eqnarray}
Therefore, the fluctuation of the gauge field $\delta A_i$ in the presence of the 
stationary flow is given as
\begin{eqnarray}
\delta A_i&=&\langle A_i^2 \rangle_{\mathsf{j}}-\langle A_i \rangle_{\mathsf{j}}^2\nonumber\\
&=& I(P)+ 2\Big\langle D_{ij}^{-1}(\vec{x})v_j(\vec{x})\partial_{x_i}(\ln P)\Big\rangle
+\Delta V_i\label{eqn:dA},
\end{eqnarray}
where $\Delta V_i=\langle D_{ji}^{-2}(\vec{x})v_j^2(\vec{x})\rangle-\langle 
D_{ji}^{-1}(\vec{x})v_j(\vec{x})\rangle^2$ 
is the fluctuation associated with the kinetic flux. We find that in addition to 
the Fisher information of the first term of Eq.(\ref{eqn:dA}) the two kinetic factors 
contribute to the fluctuation. Especially, the second term expresses interplay between 
information and flux. When the diffusion tensor is isotropic and homogeneous i.e., 
$D_{ij}(\vec{x})=D\delta_{ij}$, we have a somewhat transparent meaning of the gauge 
fluctuation. Since 
$\langle v_i(\vec{x}) \partial_{x_i} \ln P\rangle=\int \mathsf{j}_i(\vec{x}) 
\partial_{x_i}\ln Pd\vec{x}$, from the definition and the integration by parts with 
$\mathsf{j}_i(\vec{x})\ln P=0$ at the 
boundary, we have 
\begin{eqnarray}
\langle v_i(\vec{x}) \partial_{x_i} \ln P\rangle &=& 
-\int \partial_{x_i}\cdot \mathsf{j}_i(\vec{x})\ln P d\vec{x}\nonumber\\
&=& \int (\partial_t P) \ln P d\vec{x}\nonumber\\
&=& \frac{d}{dt}\int P\ln P d\vec{x}.
\end{eqnarray}
Then the second term of Eq.(\ref{eqn:dA}) corresponds to the entropy production as 
indicated \cite{Feng}
\begin{eqnarray}
-D^{-1}\langle v_i(\vec{x})\partial_{x_i}(\ln P)\rangle=\frac{dS}{dt}.
\end{eqnarray}
The third term simply reduces to the $ith$ component of the velocity fluctuation 
$\Delta V_i=D^{-2}(\langle v_i\rangle-\langle v_i\rangle^2)$. 
Therefore, the fluctuation of the gauge field is decomposed into three components; 
Fisher information, entropy production and fluctuation of the probability flux. 
\section{Fisher information matrix with covariant derivative}
In this section, we consider the effect of the gauge field on the Fisher information 
matrix. Recall that the entropy production for systems that follow a heat equation 
gives the Fisher information. This is known as the de Bruijn identity, which is one of 
the fundamental relations in information theory \cite{Cover}. It usually encapsulated 
as the statement that the time derivative of Shannon entropy equals to the Fisher 
information. We note also that this identity has an extended form for arbitrary 
nonequilibrium processes specified by the probability currents \cite{TY2}. 
Since the Fisher information $I(P)$ has the clear meaning of the dispersion of the 
particle velocity when the system follows the Fick's first 
law $\mathsf{j}_i=-D\partial_{x_i}P$
\begin{eqnarray}
I_i(P)\equiv \Big\langle \left(\frac{\partial_{x_i} P}{P}\right)^2\Big\rangle=
D^{-2}\langle v_i^2\rangle,
\end{eqnarray}
we can anticipate a legitimate counterpart of the Fisher information matrix 
corresponding to the covariant version of the Fick's first law Eq.(\ref{eqn:gnFPE}).
In this regard, we define the following quantity as a somewhat formal exploration 
of the analogue of the Fisher information matrix,
\begin{eqnarray}
I_{ij}^{cov}=\Big\langle \frac{\nabla_i P}{P}
\frac{\nabla_j P}{P}\Big\rangle,
\end{eqnarray}
where the brackets denote taking average with the probability distribution 
$P=P(\vec{x},t)$. As we shall see, this form alters the Fisher information matrix 
by replacing partial derivatives with covariant derivatives. In the following, we 
call it the covariant version of the Fisher information matrix. When we substitute 
the covariant form of the Fick's first law and expressing it by the velocity, we 
find that this tensor has an interpretation of velocity correlation 
\begin{eqnarray}
I_{ij}^{cov}(\vec{x})=\Big\langle D_{ij}^{-1}(\vec{x})v_i(\vec{x})
D_{ji}^{-1}(\vec{x})v_j(\vec{x})\Big\rangle.
\end{eqnarray}
We note that the above definition is not local gauge invariant under the gauge 
transformation $A_i\mapsto A_i+\partial_i\phi(\vec{x})$, where $\partial_i$ denotes 
$\partial/\partial x_i$ and $\phi(\vec{x})$ is a scalar function :
\begin{eqnarray}
I_{ij}^{cov}\mapsto I_{ij}^{cov}+\Big\langle \frac{\nabla_jP}{P}\partial_i\phi\Big\rangle
+\Big\langle \frac{\nabla_iP}{P}\partial_j\phi\Big\rangle
+\Big\langle (\partial_i\phi)(\partial_j\phi)\Big\rangle.
\end{eqnarray}
When the global gauge transformation $\phi(\vec{x})=const.$ applies, it keeps the gauge 
invariance. It is well known that the gauge invariance plays a profound role in theoretical 
physics, however whether or not an information quantity should satisfy the gauge principle 
has not explicitly considered. \\

In the following, let us regard that the probability distribution has a set of parameter 
$\lambda=\{\lambda_i\}$, so that the usual Fisher information matrix $I_{ij}$ 
is defined as 
\begin{eqnarray}
I_{ij}(\lambda)=\Big\langle \frac{\partial \ln P}{\partial \lambda_i}
\frac{\partial \ln P}{\partial \lambda_j}\Big\rangle,
\end{eqnarray}
where the brackets means averaging with the probability distribution 
$P_\lambda(\vec{x},t)$. In the previous section, to define an analogue of the Fisher 
information matrix we replaced the partial derivative with the covariant one. There 
still exists a possibility. An alternative for the covariant derivative is to use 
$\partial/\partial \lambda_i$ instead of $\partial/\partial x_i$, i.e., 
$\nabla_i:=\partial/\partial \lambda_i+A_i$. When there is no gauge field, the 
covariant version of the Fisher information matrix $I_{ij}^{cov}$ returns to the 
usual one $I_{ij}(\lambda)$. For stationary state (ss), we have
\begin{eqnarray}
\frac{\nabla_i P^{ss}}{P^{ss}}=\frac{\partial \ln P^{ss}}
{\partial \lambda_i}+A_i^{ss}.\label{eqn:covd}
\end{eqnarray}
Then, the covariant version of the Fisher information matrix reads,
\begin{eqnarray}
I^{cov}_{ij}(\lambda)=\Big\langle \left( \frac{\partial \ln P^{ss}}{\partial \lambda_i}
+A_i^{ss}\right)\left( \frac{\partial \ln P^{ss}}{\partial \lambda_j}+
A_j^{ss} \right)\Big\rangle,
\end{eqnarray}
where the average is taken with $P^{ss}$. Let us further assume that the stationary 
states can be characterized by the canonical equilibrium distribution with inverse 
temperature $\beta$ as,
\begin{eqnarray}
P^{ss}_\lambda(\vec{x})=e^{\beta(F(\lambda)-E(\vec{x},\lambda))},\label{eqn:cano}
\end{eqnarray}
where $F(\lambda)=-\beta^{-1}\ln\int \exp{[-\beta E(\vec{x},\lambda)]}d\vec{x}$ 
is the free energy determined by the energy of the system $E(\vec{x},\lambda)$. It is 
easily checked that from the definition of the free energy the average of the gradient 
of energy with respect to the parameter $\lambda_i$ equals to the free energy change 
i.e., $dF(\lambda)/d\lambda_i=\langle\partial E/\partial \lambda_i\rangle$ \cite{Crooks}. 
Therefore, the first term of Eq.(\ref{eqn:covd}) becomes 
\begin{eqnarray}
\frac{\partial \ln P_{\lambda}^{ss}(\vec{x})}{\partial \lambda_i}=
\beta\left( \Big\langle \frac{\partial E}{\partial \lambda_i}
\Big\rangle -\frac{\partial E}{\partial \lambda_i}\right)
\equiv \beta\Delta_{\lambda_i}E.
\end{eqnarray}
Substituting this into the definition, we have 
\begin{eqnarray}
I_{ij}^{cov}(P^{ss})&=&\Big\langle (\beta\Delta_{\lambda_i}E+A_i^{ss})
(\beta\Delta_{\lambda_j}E+A_j^{ss})\Big\rangle \nonumber\\
&=& I_{ij}(P^{ss})+(\rm{additional \ terms}),
\end{eqnarray}
where the first term is the usual Fisher information matrix for the stationary state
\begin{eqnarray}
I_{ij}(P^{ss})&=& \Big\langle \frac{\partial \ln P^{ss}}{\partial \lambda_i}
\frac{\partial \ln P^{ss}}{\partial \lambda_j}\Big\rangle\nonumber\\
&=&\beta^2\langle (\Delta_{\lambda_i}E)(\Delta_{\lambda_j}E)\rangle.
\end{eqnarray}
The additional terms are relevant to the energy of the system. 
Since $A_i^{ss}=-\partial_{x_i}\ln P^{ss}$ from Eq.(\ref{eqn:gnFPE}) for the 
stationary state, we can calculate one of the additional terms as 
\begin{eqnarray}
\langle A_i^{ss}\Delta_{\lambda_j}E\rangle&=&-\int \frac{\partial P^{ss}}{\partial x_i}
\left( \Big\langle \frac{\partial E}{\partial \lambda_j}\Big\rangle-
\frac{\partial E}{\partial \lambda_{j}} \right)d\vec{x}\nonumber\\
&=& -\Big\langle \frac{\partial^2 E}{\partial x_i\partial \lambda_{j}}\Big\rangle,
\end{eqnarray}
where we have assumed $P^{ss}\Delta_{\lambda_j}E=0$ at the boundary when the 
integration by parts is performed. This assumption is natural since there is no 
particle there. Furthermore, from the canonical form of the stationary state 
Eq.(\ref{eqn:cano}) we have 
\begin{eqnarray}
\frac{\partial P^{ss}}{\partial x_i}=-\beta \frac{\partial E}{\partial x_i}  P^{ss}.
\end{eqnarray}
Then the correlation function $\langle A_i^{ss}A_j^{ss}\rangle$ becomes
\begin{eqnarray}
\langle A_i^{ss}A_j^{ss}\rangle=\beta^2 \Big\langle \frac{\partial E}{\partial x_i}
\frac{\partial E}{\partial x_j}\Big\rangle.
\end{eqnarray}
With all additional terms considered, we have 
\begin{eqnarray}
I_{ij}^{cov}(P^{ss})=I_{ij}(P^{ss})-\beta \left(\Big\langle \frac{\partial^2 E}
{\partial x_i\partial \lambda_j}\Big\rangle+\Big\langle\frac{\partial^2 E}
{\partial x_j\partial \lambda_i}\Big\rangle\right)+
\beta^2 \Big\langle \frac{\partial E}{\partial x_i}
\frac{\partial E}{\partial x_j}\Big\rangle.
\end{eqnarray}
In this definition, we see that the deviation from the usual Fisher information 
matrix is due to the finite temperatre. It also indicates that these contributions 
become less effective at higher temperatures. 

\section{Conclusion}
The gauge field corresponding to nonequilibrium processes represented by the 
general nonlinear FPE was studied. We showed that the fluctuation of the field 
can be quantified by the Fisher information. We have introduced two types of 
covariant version of the Fisher information matrix in the settings. While the 
usual Fisher information is relevant to the fluctuation of energy of the system, 
the new definition has velocity correlation when the diffusion constant is uniform 
over the domain and symmetric. The physical interpretation of the gauge field turns 
out to be much richer and should be more important than formal formulation. 

\section*{Acknowledgments}
The author thanks for the facilities of the Shirosato-Kozukue communitty center, 
where the part of this work was performed. Stimulating discussions on covariant 
derivatives with T. Ootsuka at the Ochanomizu University is also acknowledged.

\section{Appendix} 
For an $n$-dimensional vector $\vec{v}=(v^1(\vec{x}),\ldots,v^n(\vec{x}))$,  
the covariant derivative is defined as 
\begin{equation}
\nabla_j v^i=\frac{\partial v^i}{\partial x_j}+\omega^i_{jk}v^k,\nonumber
\end{equation}
where $\omega^i_{jk}$ is the affine metric connection. 
When we deal with probability distribution functions, we only consider $i=1$ (i.e., a 
scalar function, and also an $n$-form in the $n$-dimensional space). Then, we have 
\begin{eqnarray}
\nabla_j v^1=\frac{\partial v^1}{\partial x_j}+\omega^1_{j1}v^1.\nonumber
\end{eqnarray}
If we regard the coefficients of the affine connection $\omega^1_{j1}$ as the $j$th 
component of the gauge field $A_j$, then a probability distribution function $P=v^1$
satisfies 
\begin{eqnarray}
\frac{\nabla_jP}{P}=\partial_j\ln P+A_j.\nonumber
\end{eqnarray}

\end{document}